\documentstyle[onecolumn,epsfig]{mn}
\newif\ifAMStwofonts
\ifoldfss
  \ifCUPmtlplainloaded \else
    \NewTextAlphabet{textbfit} {cmbxti10} {}
    \NewTextAlphabet{textbfss} {cmssbx10} {}
    \NewMathAlphabet{mathbfit} {cmbxti10} {} % for math mode
    \NewMathAlphabet{mathbfss} {cmssbx10} {} %  "   "    "
  \fi
  \ifAMStwofonts
    \ifCUPmtlplainloaded \else
      \NewSymbolFont{upmath} {eurm10}
      \NewSymbolFont{AMSa} {msam10}
      \NewMathSymbol{\upi}     {0}{upmath}{19}
      \NewMathSymbol{\umu}     {0}{upmath}{16}
      \NewMathSymbol{\upartial}{0}{upmath}{40}
      \NewMathSymbol{\leqslant}{3}{AMSa}{36}
      \NewMathSymbol{\geqslant}{3}{AMSa}{3E}

      \let\leq=\leqslant 
      \let\geq=\geqslant 
    \fi
  \fi
\fi % End of OFSS

\ifnfssone
  \newmathalphabet{\mathit}
  \addtoversion{normal}{\mathit}{cmr}{m}{it}
  \addtoversion{bold}{\mathit}{cmr}{bx}{it}
  \newmathalphabet{\mathbfit} % math mode version of \textbfit{..}
  \addtoversion{normal}{\mathbfit}{cmr}{bx}{it}
  \addtoversion{bold}{\mathbfit}{cmr}{bx}{it}
  \newmathalphabet{\mathbfss} % math mode version of \textbfss{..}
  \addtoversion{normal}{\mathbfss}{cmss}{bx}{n}
  \addtoversion{bold}{\mathbfss}{cmss}{bx}{n}
  \ifAMStwofonts
    \ifCUPmtlplainloaded \else
      \UseAMStwoboldmath
      \makeatletter
      \new@mathgroup\upmath@group
      \define@mathgroup\mv@normal\upmath@group{eur}{m}{n}
      \define@mathgroup\mv@bold\upmath@group{eur}{b}{n}
      \edef\UPM{\hexnumber\upmath@group}
      \new@mathgroup\amsa@group
      \define@mathgroup\mv@normal\amsa@group{msa}{m}{n}
      \define@mathgroup\mv@bold\amsa@group{msa}{m}{n}
      \edef\AMSa{\hexnumber\amsa@group}
      \makeatother
      \mathchardef\upi="0\UPM19
      \mathchardef\umu="0\UPM16
      \mathchardef\upartial="0\UPM40
      \mathchardef\leqslant="3\AMSa36
      \mathchardef\geqslant="3\AMSa3E

      \let\leq=\leqslant 
      \let\geq=\geqslant 
    \fi
  \fi
\fi % End of NFSS release 1

\ifnfsstwo
  \DeclareMathAlphabet{\mathbfit}{OT1}{cmr}{bx}{it}
  \SetMathAlphabet\mathbfit{bold}{OT1}{cmr}{bx}{it}
  \DeclareMathAlphabet{\mathbfss}{OT1}{cmss}{bx}{n}
  \SetMathAlphabet\mathbfss{bold}{OT1}{cmss}{bx}{n}
  \ifAMStwofonts
    \ifCUPmtlplainloaded \else
      \DeclareSymbolFont{UPM}{U}{eur}{m}{n}
      \SetSymbolFont{UPM}{bold}{U}{eur}{b}{n}
      \DeclareSymbolFont{AMSa}{U}{msa}{m}{n}
      \DeclareMathSymbol{\upi}{0}{UPM}{"19}
      \DeclareMathSymbol{\umu}{0}{UPM}{"16}
      \DeclareMathSymbol{\upartial}{0}{UPM}{"40}
      \DeclareMathSymbol{\leqslant}{3}{AMSa}{"36}
      \DeclareMathSymbol{\geqslant}{3}{AMSa}{"3E}

      \let\leq=\leqslant 
      \let\geq=\geqslant 
    \fi
  \fi
\fi % End of NFSS release 2

\ifCUPmtlplainloaded \else
  \ifAMStwofonts \else % If no AMS fonts
    \def\upi{\pi}
    \def\umu{\mu}
    \def\upartial{\partial}
  \fi
\fi

\title[Iron K$\alpha$ line profiles driven by non-axisymmetric 
illumination]{Iron K${\bmath{\alpha}}$ line profiles driven by non-axisymmetric illumination}
\author[Q. Yu \& Y. Lu]
       {Qingjuan Yu$^{1,2}$\thanks{Current address: Peyton Hall, Princeton
University, Princeton, NJ 08544-1001, USA. E-mail address: yqj@astro.princeton.edu} and
Youjun Lu$^1$\thanks{E-mail address: lyj@astro.princeton.edu}\\
$^1$ Centre for Astrophysics, University of Science and Technology of China,
Hefei Anhui 230026, P.\ R.\ China\\
$^2$ Princeton University Observatory, Princeton, NJ 08544-1001, USA}
\date{Accepted ?.
      Received ?;
      in original form May 8, 1998}

\pagerange{\pageref{firstpage}--\pageref{lastpage}}
\pubyear{1998}

\begin{document}
\maketitle

\label{firstpage}

\begin{abstract}
Previous calculations of Fe K$\alpha$ line profiles are based on axisymmetric 
emissivity laws. In this paper, we show line profiles driven by 
non-axial symmetric illumination which results from an off-axis X-ray point
source. We find that source location and motion have significant
effects on the red wing and blue horn of the line profiles.
The disk region under the source will receive more flux, which is the most
important factor to affect the line profiles. We suggest that at least part
of the variation in Fe K$\alpha$ line profiles is caused by the motion of
X-ray sources. Future observations of Fe K$\alpha$ line profiles will
provide more information about the distribution and motion of the
X-ray sources around black holes, and hence the underlying physics.
\end{abstract}


%%%%%%%%%%%%%%%%%%%%%%%%%%%%%%%%%%%%%%%%%%%%%%%%%%%%%%%%%%%%%%%%%%%%%%%%%%%%%
\section{Introduction}
The broad Fe K$\alpha$ emission line at 6.4keV in many Seyfert 1 galaxies 
is a powerful probe of the environment around
a black hole. Though its physical mechanism has been interpreted in many ways, 
recent observations by {\it ASCA} show that the asymmetric profile of Fe 
K$\alpha$ emission line is much better explained by gravitational broadening, 
Doppler effect and accretion-disk origin (Tanaka et al.\
1995; Fabian et al.\ 1995; Nandra et al.\ 1997; Sulentic et al.\
1997 and references therein).
Thus, Fe K$\alpha$ line profiles provide a valuable clue to explore the
information about the gravitational field, the accretion disk structure,
the X-ray continuum source geometry, the inclination between an observer
and accretion disk axes and the evidence for the existence of a black hole.
Fe K$\alpha$ line profile modeling has attracted much
attention and the factors above should be involved in any model
based on accretion-disk origin. 
Theoretical Fe K$\alpha$ line profiles produced in the 
Schwarzschild metric were first fully relativistically treated by Fabian
et al.\ (1989).
Laor (1991) and  Kojima (1991) extended those to the Kerr metric.
They grasped the two most important physical mechanisms---Doppler effect and
gravitational redshift---to produce the Fe K$\alpha$ line profiles,
and discussed the important influence of inclination. The detailed
structures of the accretion disk/X-ray sources were, however, neglected.
Their calculations were
formulated on some very simple assumptions, such as the axisymmetric
power-law line emissivity law and the Keplerian motion of disk material.

The differences in the line profiles from source to source
have implied variations in the geometry and structure of accretion-disk/X-ray
source, which cannot be accounted for solely by inclination or the types of
central black holes (Nandra et al.\ 1997). In order to 
adequately model the Fe K$\alpha$ emission, many detailed calculations about
the accretion disk must be performed (Krolik 1996). Considering the  
geometry (flat, conical or concave),
size and ionization state of the accretion disk, some authors have derived
more realistic emissivity law by performing Monte Carlo simulations of Compton
reflection from a neutral or ionized medium, including fluorescent iron 
emission (George \& Fabian 1991; Matt, Perola \& Piro 1991; Matt et al.\ 1992;
Matt, Fabian \& Ross 1993b; \.{Z}ycki \& Czerny 1994).
A more complete calculation was provided by Matt, Fabian \& Ross (1996),
in which the
ionization balance is treated as a function of depth and local angular
distribution, and Compton scattering and resonant absorption are taken into
account in detail. Some improvement on the 
kinematics of the accretion disk has also been incorporated into the
calculation of Fe K$\alpha$ line profiles. The emitting
materials are no longer limited to orbits in the equatorial plane
of a black hole. The effects of the finite thickness, radial accretion flow
and turbulence were considered (Pariev \& Bromley 1997). Rybicki \& Bromley
(1997) calculated line formation in the presence of velocity gradients induced
by the differential flow in the disk. Recently, Usui, Nishida \& Eriguchi
(1998) show the Fe
K$\alpha$ line profiles from self-gravitating disks or toroids which play an 
important role for the gravitational field, the dynamics of the disk inner
edge and the disk structure.

To date, many authors have incorporated the structure of disks into the
calculation of Fe K$\alpha$ lines, but another important factor affecting line
profiles---the geometry of X-ray sources, is generally left in an idealized state. The
X-ray source is mostly assumed to be a static point located on the disk axis.
Matt et al.\ (1991) extended the source to be a uniform optically thin
corona or a homogeneous
optically thick sphere to investigate its effects on equivalent width. In that
extended-source geometry, the ionization parameter spans a much narrower range
of values along disk radii than in the point-source geometry. 
Matt et al.\ (1993b) discussed its influences on line profiles,
but the disk illumination is still confined to be axisymmetric just as the
illumination of an axial point source.

In the past several years, {\it Ginga} and {\it GRO} OSSE observation have
led to much progress in X-ray source modeling.
(Gondek et al.\ 1996; Zdziarsiki et al.\ 1995).
Exponentially cut-off power laws of RQ Seyfert 1 X-ray spectra can be modeled
by comptonization models either in a non-thermal plasma (Zdziarski, Coppi
\& Lamb
1990) or in a thermal hot corona covering the disk (Haardt \& Maraschi 1991,
1993; Field \& Rogers 1993; Zdziarski et al. 1994). Haardt \& Maraschi (1991, 1993)
presented a uniform, plane-parallel corona
and extended it to a patchy structured corona model (Haardt, Maraschi \&
Ghisellini 1994), since the
UV fluxes in many Seyfert 1s are much larger than the X-ray fluxes (Walter \&
Fink 1993). They
proposed that the X-ray emission from radio-quiet active galactic nuclei and
galactic black holes is due to comptonization of soft thermal photons emitted 
by the underlying accretion disk in localized structures. Stern et al.\ (1995)
provided more arguments for the idea that the X-rays come from a
number of individual active regions located above the surface of the disk.
If so, the flux striking the disk will be determined by 
the distribution of the blobs and the relativistic motion between the source and
the accretion disk. The illumination is not uniform and it depends not only on
disk radii but also on polar angles of the disk plane, which will affect the
Fe K$\alpha$ line profile because the different energy parts of the Fe
K$\alpha$ line profile stem from different angular and radial regions of an
accretion disk due to Doppler effect and gravitational broadening
(e.g.\ Matt, Perola \& Stella 1993a).

In this paper, we will focus on how the location and motion of one off-axis
X-ray source affect the time-averaged Fe K$\alpha$ line profile. We consider an
accretion disk around a Schwarzschild black hole. We assume that the disk
is illuminated only by one off-axis point source. The flux striking the
accretion disk, which is non-axisymmetric, drives the iron line fluorescence. 
Our results will show the information about the distribution and
motion of the X-ray sources can be extracted from the Fe K$\alpha$ line profiles.
%%%%%%%%%%%%%%%%%%%%%%%%%%%%%%%%%%%%%%%%%%%%%%%%%%%%%%%%%%%%%%%%%%%%%%%%%%%%%%
\section{Disk-line in the Schwarzschild metric}
In the Schwarzschild metric, the line element is written as follows:
\begin{equation}
{\rm d}s^{2}=-(1-2/r){\rm d}t^{2}+\frac{1}{1-2/r}{\rm d}r^{2}+
r^{2}{\rm d\theta}^{2}+
r^{2}{\rm sin}^{2}\theta{\rm d}\varphi^{2}.
\end{equation}
where we use the units of $G=c=M=1$ ($M$ is the black hole mass).

The observer is assumed to be located at ($r_{\rm obs},\theta_{\rm obs},
\varphi_{\rm obs}$) where $r_{\rm obs}=1000$ and
$\varphi_{\rm obs}=0^{\circ}$. We set the X-ray point source at $(r_{\rm s},\theta_{\rm s},
\varphi_{\rm s})$. We define frames
locally moving with the disk materials as $R^{\ast}$ (see Appendix).
All terms computed in those frames
are labeled with a star.
%%%%%%%%%%%%%%%%%%%%%%%%%%%%%%%%%%%%%%%%%%%%%%%%%%%%%%%%%%%%%%%%%%%%%%%%%%%%%%
\subsection{Photon motions in the Schwarzschild metric}
The photons follow null geodesics either between the disk and the X-ray source
or between the disk/X-ray source and the observer (Carter 1968; Misner, Thorne
\& Wheeler 1973).
In the Schwarzschild metric, the expressions are simplified as:
\begin{equation}
r^{2}\frac{{\rm d}r}{{\rm d}\lambda}=\pm EV_{\rm r}^{1/2}
\end{equation}
\begin{equation}
r^{2}\frac{{\rm d}\theta}{{\rm d}\lambda}=\pm EV_{\rm\theta}^{1/2}
\end{equation}
\begin{equation}
r^{2}\frac{{\rm d}\varphi}{{\rm d}\lambda}=\frac{El}{{\rm sin}^{2}\theta}
\end{equation}
\begin{equation}
r^{2}\frac{{\rm d}t}{{\rm d}\lambda}=\frac{Er^{4}}{r^{2}-2r}
\end{equation}
\begin{equation}
V_{\rm r}=(r^{2})^{2}-(r^{2}-2r)(l^{2}+q^{2})
\end{equation}
\begin{equation}
V_{\rm\theta}=q^{2}-l^{2}{\rm ctg}^{2}\theta
\end{equation}
where E, q, l are motion constants and $\lambda$ is an affine parameter related to the
proper time.
 
Then, the photon motion equations are given by:
\begin{equation}
\int\frac{{\rm d}r}{V_{\rm r}^{1/2}}=\pm\int\frac{{\rm d}\theta}{V_{\rm\theta}^{1/2}};
\int {\rm d}\varphi=\pm\int\frac{l{\rm d}\theta}{{\rm sin}^{2}\theta V_{\rm\theta}^{1/2}}.
\end{equation}
%%%%%%%%%%%%%%%%%%%%%%%%%%%%%%%%%%%%%%%%%%%%%%%%%%%%%%%%%%%%%%%%%%%%%%%%%%%%%%
\subsection{The observed line flux}
Due to relativistic effects, photons emitted with the
rest frequency $\nu_{\rm em}^{\ast}$ will reach the observer with a frequency
$\nu_{\rm obs}$, their ratio being given by: 
\begin{equation}
g=\frac{\nu_{\rm obs}}{\nu_{\rm em}^{\ast}}=
\frac{\bmath{p_{\rm obs}\cdot u_{\rm obs}}}{\bmath{p_{\rm em}\cdot u_{\rm em}}}.
\end{equation}

We do not consider the effect of higher order images of the disk (Bao, Hadrava \& \O stgaard 1994a,b).
The observed flux distribution $F(E_{\rm obs})$ at the observed energy is
given by (Cunningham 1975):
\begin{equation}
{\rm d}F(E_{\rm obs})=I_{\rm obs}(E_{\rm obs}){\rm d}\Omega
=g^{3}I_{\rm em}^{\ast}(E_{\rm em}^{\ast}){\rm d}\Omega
\end{equation}
where $d\Omega$ is the solid angle subtended by the disk in the observer's sky
and we have made use of the relativistic invariance of $I_{\nu}/\nu^{3}$, 
$I_{\nu}$ being the specific intensity.

Two important ingredients involved in determining the Fe K$\alpha$
emissivity are the structure of the disk and the number of illuminating photons
with energy greater than the photoelectric threshold for neutral iron.
 In this paper, our assumptions about the disk are the same as those in the
paper by Reynolds, Young \& Begelman et al. (1999). The disk region emitting
the iron
fluorescence is assumed to be between $r_{\rm in}=2$ (the horizon) and
$r_{\rm out}=200$. Outside the marginally stable orbit $r_{\rm ms}$ 
($r_{\rm ms}\leq r_{\rm em}\leq r_{\rm out}$), monochromatic 6.4keV cold
Fe K$\alpha$ line is isotropically (in the frame moving with the disk)
emitted from the geometrically thin and optically thick disk, where the disk
material is essentially in circular, Keplerian motion. 
Inside the radius of marginal stability, the disk material is assumed to be
in free-fall and the line emission is dependent on the ionization
parameter $\xi$. For $\xi\leq 100$ ergs cm ${\rm s^{-1}}$, we assume a cold iron
fluorescence line at 6.4keV with the strength given by the 
standard neutral slab calculations. For $\xi$ in the range of 100-500
ergs cm ${\rm s^{-1}}$ we assume no line emission because of the efficient
resonance trapping and Auger destruction of these line photons. For $\xi$ in
the range 500-5000 ergs cm ${\rm s^{-1}}$ we assume a blend of helium-like and
hydrogen-like 
iron line emission with rest frame energies of 6.67 and 6.97 keV, respectively,
with an effective fluorescent yield for each line the same as that for the 
neutral case. For $\xi\geq$ 5000 ergs cm ${\rm s^{-1}}$, the material is taken
to be completely ionized, and no line emission results.

Thus, the specific intensity in the rest frame
$I_{\rm em}^{\ast}(E_{\rm em}^{\ast})$ is given as follows:
\begin{equation}
I_{\rm em}^{\ast}(E_{\rm em}^{\ast})=\epsilon^{\ast}(r_{\rm em},\varphi_{\rm em})
\delta(E_{\rm em}^{\ast}-E_{0}),\ \ \ \ (r_{\rm in}\leq r_{\rm em}\leq r_{\rm out}),
\end{equation}
where $\epsilon^{\ast}(r_{\rm em},\varphi_{\rm em})$ is the emissivity law
and $E_{0}=6.4$, 6.67 or 6.97keV depending on $r_{\rm em}$ and $\xi$. The total
observed flux distribution is given by
\begin{equation}
F(E_{\rm obs})=\int \epsilon^{\ast}(r_{\rm em},\varphi_{\rm em})g^{4}
\delta(E_{\rm obs}-gE_{0}){\rm d}\Omega.
\end{equation}

The emission is assumed to be isotropic in the source rest frame.
In the frame moving with the disk materials, the illumination flux,
which is emitted from the X-ray source, is denoted by $F^{*}(r,\varphi)$.
We shall refer to $F^{*}(r,\varphi)$ as the illumination
law. The line enhancement due to the returning radiation (Cunningham 1976;
Dabrowski et al.\ 1997) is neglected. For the matter at a given ionization state,
the iron fluorescent emissivity
will be proportional to the illumination law, $\epsilon^{\ast}(r,\varphi)\propto
F^{\ast}(r,\varphi)$.
%%%%%%%%%%%%%%%%%%%%%%%%%%%%%%%%%%%%%%%%%%%%%%%%%%%%%%%%%%%%%%%%%%%%%%%%%%%%%%
\section{Calculations and Results}
%%%%%%%%%%%%%%%%%%%%%%%%%%%%%%%%%%%%%%%%%%%%%%%%%%%%%%%%%%%%%%%%%%%%%%%%%%%%%%
\subsection{Non-axisymmetric illumination law}
For a X-ray point source on the disk axis or a uniform slab above the disk, the
illumination
law is generally simplified as axially symmetric, such as 
$r_{\rm em}^{q}$, $H/(r_{\rm em}^{2}+H^{2})^{3/2}$ or
$H/(r_{\rm em}^{2}+H^{2})^{3/2}(1-2/r_{\rm em})^{3+\alpha}$ (Reynolds \&
Begelman 1997), where $H$ is the X-ray
source height above the disk and $\alpha$ is the energy index of the source
spectrum $F(\nu)\propto\nu^{-\alpha}$. If the source deviates from the 
axis, the symmetry will be violated. First, in the Newtonian
case, the illumination will become symmetric about the axis jointing the 
source with the point on the disk which is just under the source, not 
symmetric about the disk axis. Disk regions closer to the source receive more
flux ,
 which will be the most important factor resulting in the non-axisymmetric
illumination law. Second, at a given radius, the relative motion of the
source and each angular region of the disk is no longer identical, which also 
cause the asymmetry of the illumination law in the frame moving with the
disk materials. In addition, the gravitational lensing effect cannot be neglected.
More photons focus towards the black hole. The number of the photons arriving
within the range $r_{\rm em}$---$r_{\rm em}+{\rm d}r_{\rm em}$ and
$\varphi_{\rm em}$---$\varphi_{\rm em}+{\rm d}\varphi_{\rm em}$
on the disk will depend not only on $r_{\rm em}$ but also on $\varphi_{\rm em}$
since the source is off axis.
In this paper, using the photon motion equation(8), we take into account
both the general and special relativistic effect 
to work out the time-averaged illumination law (see Appendix A for details).
In the calculation, for different source locations and motions, we assume
the same X-ray emitting power in the source rest frames, without considering
the time differences of the photon motions from the source to the disk.

\begin{figure}
\epsfig{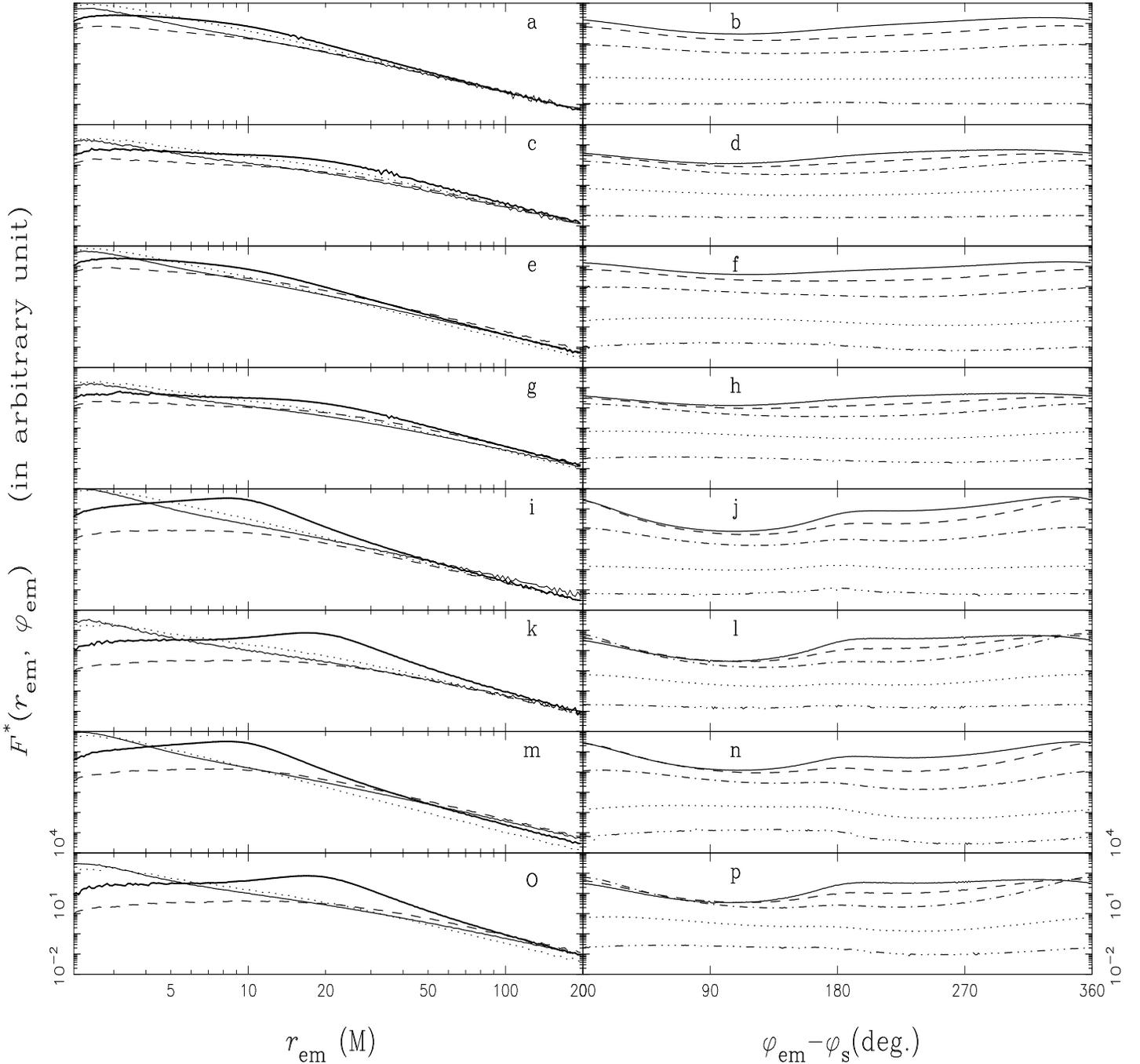}
\vskip 0.15cm
\caption{The illumination laws produced by the X-ray point source with different
locations and angular velocities (see Appendix):
$r_{\rm s}=10, \theta_{\rm s}=30^{\circ}, \varpi_{\rm s}=0$ (a,b);
$r_{\rm s}=20, \theta_{\rm s}=30^{\circ}, \varpi_{\rm s}=0$ (c,d);
$r_{\rm s}=10, \theta_{\rm s}=30^{\circ}, \varpi_{\rm s}=\frac{1}{2}\varpi_{\rm em}(r_{\rm s})$
(e,f);
$r_{\rm s}=20, \theta_{\rm s}=30^{\circ}, \varpi_{\rm s}=\frac{1}{2}\varpi_{\rm em}(r_{\rm s})$
(g,h);
$r_{\rm s}=10, \theta_{\rm s}=60^{\circ}, \varpi_{\rm s}=0$ (i,j);
$r_{\rm s}=20, \theta_{\rm s}=60^{\circ}, \varpi_{\rm s}=0$ (k,l);
$r_{\rm s}=10, \theta_{\rm s}=60^{\circ}, \varpi_{\rm s}=\frac{1}{2}\varpi_{\rm em}(r_{\rm s})$
(m,n);
$r_{\rm s}=20, \theta_{\rm s}=60^{\circ}, \varpi_{\rm s}=\frac{1}{2}\varpi_{\rm em}(r_{\rm s})$
(o,p).
Left column: $\varphi_{\rm em}-\varphi_{\rm s}=0^{\circ}$ (thick solid line),
$\varphi_{\rm em}-\varphi_{\rm s}=90^{\circ}$ (dashed line),
$\varphi_{\rm em}-\varphi_{\rm s}=180^{\circ}$ (thin solid line),
$\varphi_{\rm em}-\varphi_{\rm s}=270^{\circ}$ (dotted line);
right column: $r_{\rm em}=6.2$ (solid line), $r_{\rm em}=10.1$ (dashed line),
$r_{\rm em}=20.2$ (dot-dashed line), $r_{\rm em}=60.0$ (dotted line),
$r_{\rm em}=151.5$ (dot-dot-dot-dashed line).
}
\end{figure}

Fig.1 demonstrates the illuminating flux in the disk material rest frame as
a function of $r_{\rm em}$ and $\varphi_{\rm em}-\varphi_{\rm s}$ for different
locations and motions of X-ray sources.
As seen from the figure, the region below the source
($\varphi_{\rm em}\approx\varphi_{\rm s}, r_{\rm em}\approx$$r_{\rm s}{\rm sin}\theta_{\rm s}$) receives more flux,
which will be the most important factor to affect the Fe K$\alpha$ line
profile. At the outer region of the disk,
the illumination laws essentially maintain power law of the radii.
Since the X-ray source is generally located at the vicinity of the black hole,
the illumination at the outer region depends more strongly on the angle
$\varphi_{\rm em}-\varphi_{\rm s}$ in the inner region than in the outer region. 
Besides, some features of relativistic effects can be found in the Fig.1.
When the source inclination is high (say, $\theta_{\rm s}=60^{\circ}$), the
gravitational lensing effect is shown in Fig.1 (a little bump at
$\varphi_{\rm em}-\varphi_{\rm s}\approx 180^{\circ}$ in Fig.1j,l,n,p).
If the source is static, the photon number striking the disk should be
symmetric about $\varphi_{\rm em}-\varphi_{\rm s}=180^{\circ}$.
Including the relativistic effects, the illumination on the receding
side to the source ($\varphi_{\rm em}-\varphi_{\rm s}\approx 90^{\circ}$) is
generally less than that on the approaching side
($\varphi_{\rm em}-\varphi_{\rm s}\approx 270^{\circ}$) due to Doppler boosting.
In additon, it is more realistic that
the source has a velocity because the corona in the black hole proximity is
generally coupled with the accretion
disk via some physical mechanism. For example, Field \& Rogers (1993) argued that the
corona is magnetically confined by loops of magnetic field which have
footprints in the accretion disk. Here, we simply suppose that the source moves
with an angular velocity
$\bmath{u_{\rm s}}=(u_{\rm s}^{t}, u_{\rm s}^{r}, u_{\rm s}^{\theta}, u_{\rm s}^{\varphi})
=(u_{\rm s}^{t}, 0, 0, u_{\rm s}^{\varphi})$ as the disk materials do
(see Appendix). At the very outer radii, that the 
illumination on the receding side becomes larger than that on the approaching 
side is just due to the relativistic motion between the source and the disk
materials (Fig.1e,g,m,o). That is, the source has a velocity towards the
receding side and away from the approaching side.
%%%%%%%%%%%%%%%%%%%%%%%%%%%%%%%%%%%%%%%%%%%%%%%%%%%%%%%%%%%%%%%%%%%%%%%%%%%%%%
\subsection{Iron K$_{\alpha}$ fluorescent line profiles}
Fanton et al.\ (1997) summarized two integration techniques (direct integration of
the geodesic equations and ray tracing) to model emission
line profiles. Choosing the ray tracing approach, we use elliptic
integrals (Rauch \& Blandford 1994)
to obtain the Fe K$\alpha$ line profiles with high efficiency. We calculate
the trajectories of photons that propagate from the sky plane at the observer
to the accretion disk. For each point in the photographic plate of the observer
--- that is for each impact parameter $(\alpha,\beta)$--- the constants of
motion $(l,q)$ are calculated using the following equations (Cunningham \&
Bardeen 1973):
\begin{equation}
\alpha=-\frac{l}{{\rm sin}\theta}, 
\beta=(q^{2}-l^{2}{\rm ctg}^{2}\theta)^{1/2}.
\end{equation}
With the equations(8), we can 
obtain the location $(r_{\rm em}, \frac{\pi}{2}, \varphi_{\rm em})$ where the
Fe K$\alpha$ line emits with motion constants $(l, q)$. 
Then by binning the arrived flux at each pixel in the disk
image, the emission line profile is obtained using the illumination at
$(r_{\rm em}, \frac{\pi}{2}, \varphi_{\rm em})$ calculated in Sect.3.1.
In this paper, the contribution to the line profile from the region inside
the marginally stable orbit is not significant since the source is not set to 
be very close to the black hole. The contribution from the region 
$r\leq r_{\rm ms}$ is significant only when the source is inefficient and
close to the black hole.
\begin{figure}
\epsfig{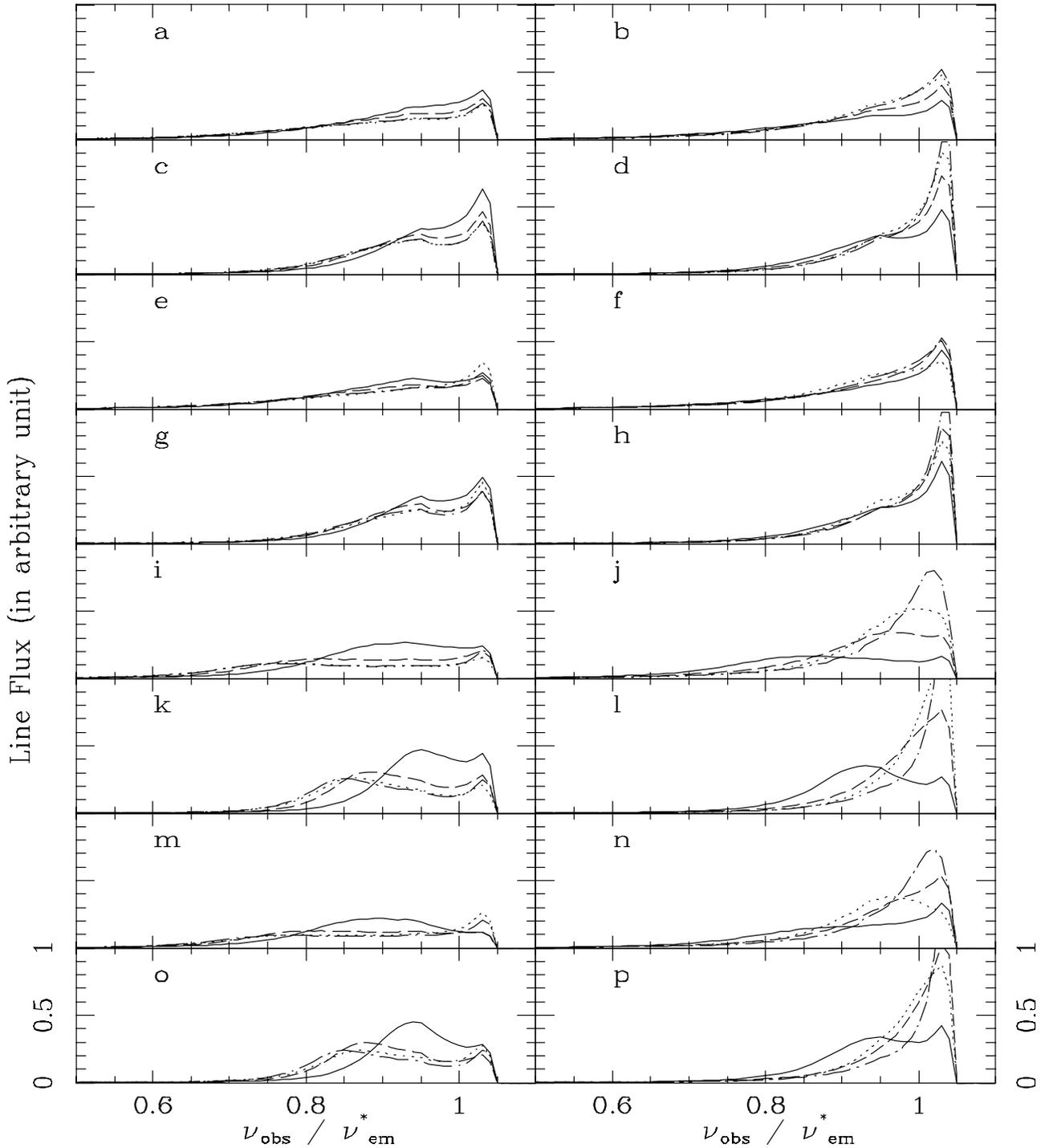}
\vskip 0.15cm
\caption{The Fe K$\alpha$ line profiles for different locations and motions
of X-ray point sources ($\nu^{*}_{\rm em}=6.4{\rm keV}, r_{\rm obs}=1000,
\varphi_{\rm obs} =0^{\circ}$).
The observer inclination $\theta_{\rm obs}$ is $30^{\circ}$. 
The parameters about the source are the same as those in Fig.1.
Left column:
$\varphi_{\rm s}=0^{\circ}$ (solid line), $\varphi_{\rm s}=45^{\circ}$ (dashed line),
$\varphi_{\rm s}=90^{\circ}$ (dot-dashed line), $\varphi_{\rm s}=135^{\circ}$ (dotted line);
right column:
$\varphi_{\rm s}=180^{\circ}$ (solid line), $\varphi_{\rm s}=225^{\circ}$ (dashed line),
$\varphi_{\rm s}=270^{\circ}$ (dot-dashed line), $\varphi_{\rm s}=315^{\circ}$ (dotted line).}
\end{figure}
\begin{figure}
\epsfig{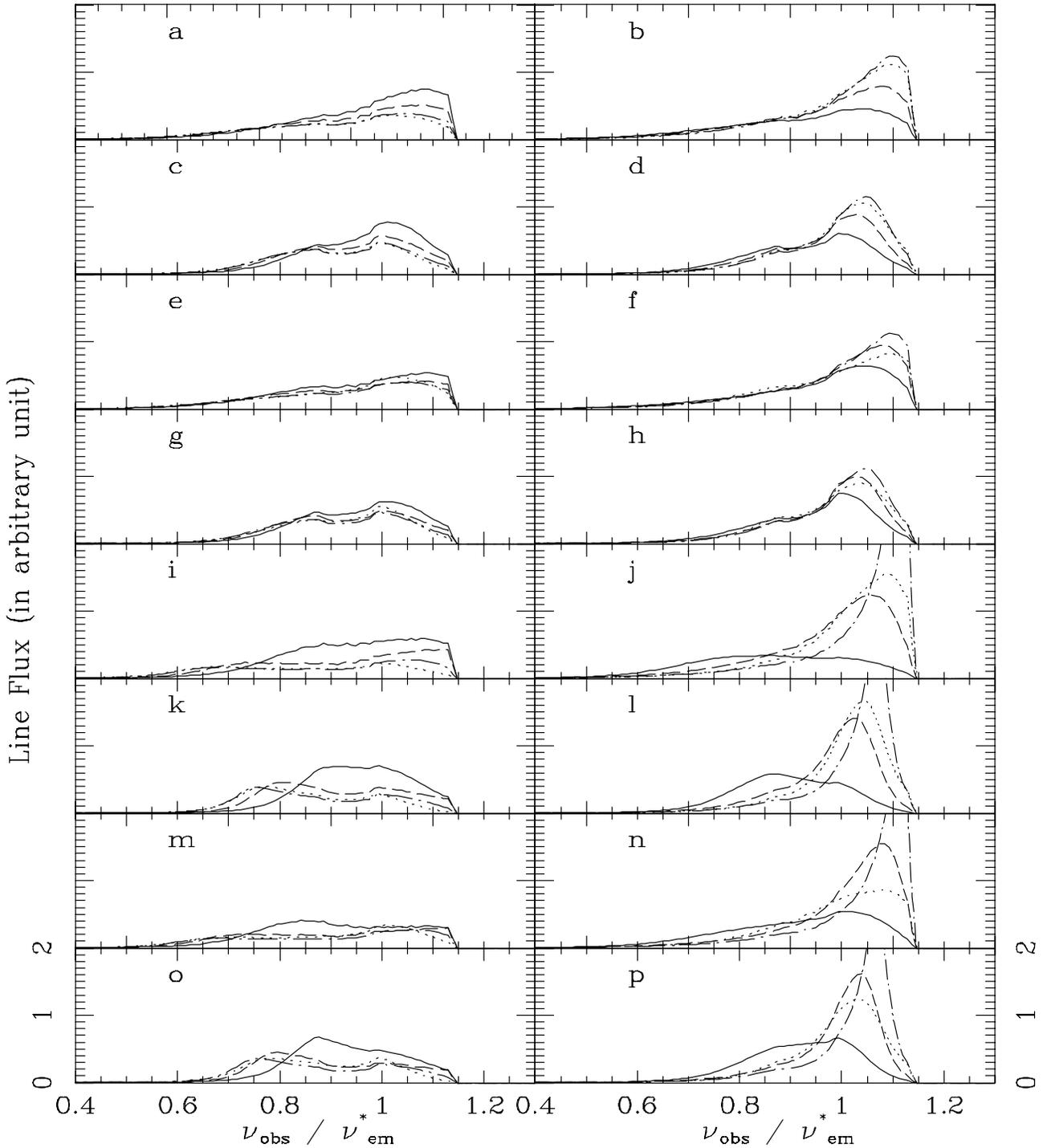}
\vskip 0.15cm
\caption{The parameters are the same as those in the Fig.2 except that the
observer inclination $\theta$ is $60^{\circ}$.}
\end{figure}

When the observer inclination is very low (i.e., the disk is face-on), at a
given radius, every angular region on the disk almost contributes to the same 
energy part of the line profile and the frequency shift of the observed photon
is approximately only a function of the disk radius $r_{\rm em}$, not
 $\varphi_{\rm em}$. 
At its corresponding disk radius, the strength of the line at each frequency is
proportional to the integration of illumination over $\varphi_{\rm em}$.
Thus, the X-ray source geometry affects the line profile simply
when the observer inclination is very low.
However, when the inclination of the observer is not too low, every part of the line
profile corresponds to a certain radial and angular region on the disk
(e.g.\ Matt et al.\ 1993a).
The blue horn comes mainly from the approaching side to the observer and the
red part stems from the receding side of the inner disk region.
The line profile is affected in a complicated way by the illumination
distribution on
two parameters $r_{\rm em}$ and $\varphi_{\rm em}$.
In this paper, we focus more on discussing that case.
From the results in Sect.3.1, the 
illumination is non-axisymmetric when the source deviates from the disk axis
and more flux arrives at the region below the source.
When the source is above the receding side of the disk
($\varphi_{\rm s}\approx 90^{\circ}$), the illumination on the approaching side
($\varphi_{\rm s}\approx 270^{\circ}$) will be less, which results in
weak blue part of the Fe K$\alpha$ line profile. Correspondingly, the red wing
of the line profile is strengthened (Fig.2-3).
Likewise, when the source is above the approaching side, the blue horn
increases and the red wing is weakened.
Those results are also consistent with the one that the two-dimensional transfer
function becomes a $\delta$ function when $\theta_{\rm s}$ is high
(Reynolds, Young \& Begelman (1999). As seen from Fig.2-3, when the source
inclination is high ($\theta_{\rm s}=60^{\circ}$), the blue horn and the red
wing both vary with $\varphi_{\rm s}$ more violently than they
do when $\theta_{\rm s}$ is low ($\theta_{\rm s}=30^{\circ}$).
The reason is that the illumination differences between
the approaching and receding side become larger when $\theta_{\rm s}$ is high.
%%%%%%%%%%%%%%%%%%%%%%%%%%%%%%%%%%%%%%%%%%%%%%%%%%%%%%%%%%%%%%%%%%%%%%%%%%%%%%
\section{Discussion}
The differences in the line profiles from source to source imply 
variations in the geometry of X-ray sources (Nandra et al. 1997).
The variability of some Seyfert 1 galaxies (e.g. MCG--6-30-15, Reynolds \& 
Begelman, 1997) is also explained by the geometry of X-ray sources.
In this paper, we explored how the time-averaged line profiles behave for
different location and motion of one off-axis source.
Generally, the observation angle mainly determines the highest energy that the
blue horn can reach.
Kojima (1991) examined the difference in the line profiles with different
inclination angle, Kerr parameter and disk parameters.
By comparing the profiles in his paper with the ones in this paper, it is found
that at a given source radius and observation angle, the most significant
factor affecting the line profiles comes from different X-ray source coordinate
$\varphi_{\rm s}$. Moreover, it is more realistic that the source has a velocity
$u_{\varphi_{\rm s}}$ with the disk materials.
The motion of the source will affect the strength of the
continuum, too. If the source is above the approaching part of the disk, the
blue peak goes up and the observed X-ray continuum is also Doppler boosted;
whereas, when the source is above the
receding part, the blue horn goes down and the continuum decreases
because the source is moving away from us. Those results are also
qualitatively true for a Kerr black hole.

There are reasons to believe that in a physically realistic situation,
there are a number of active regions or blobs in the patchy corona,
rather than one off-axis point source, which collectively drive the iron
fluorescence(Haardt, Maraschi, \& Ghisellini 1994).
The iron line profiles may provide some information about the distribution of
those X-ray blobs. The disk regions below those discrete blobs will receive
more flux to emit iron fluorescence.
It has been pointed out that the Fe K$\alpha$ line profiles of some Seyfert 
1 galaxies (say, MCG--6-30-15) are perhaps combined with different
components (Sulentic et al.\ 1998).
Here, we
suppose that the different components may correspond to the discrete blobs.
If there are many blobs in the corona, the addition of these numerous
blobs' illumination may be similar to the results from a uniform thin
corona (Matt et al. 1991). While, if there are only several blobs, the 
addition of those blobs' illumination should still be non-axial symmetric.
The flux of the blue horn will be much more correlated with the number and
the motion of the blobs above the approaching side of inner disk region;
while, the red wing will reflect the state of blobs above the receding side of
disk. 
%Some natures of the blobs may be revealed by the variation of the X-ray
%continuum and the Fe K$\alpha$ line.

Based on the calculation of the off-axis illumination, we suggest that the
location and hence the motion of the X-ray source is an alternative explanation
for the variation and reddening of the Fe K$\alpha$ line profile. Indeed, it
is more realistic that the X-ray source moves near the central black hole.
Such an explanation of Fe K$\alpha$ variability can naturely explain the
variability time-scale.
We can conclude that at least part of the variation of Fe K$\alpha$ line
is caused by the motion of the X-ray source. In order to understand the motion
of the X-ray source, we should consider both the time difference of the photon
motions from the flare to the disk and from the disk to the observer instead
of the time-averaged line profile presented here. If the source is static or
its life-time is very short(say, $<10^{2}$s, for an AGN black hole with 
$10^{7}M_{\sun}$ whose light crossing time of the gravitational radius is 
50s), the time-averaged line
profile calculated in this paper should reflect the observed time-averaged line
profile, which is the time accumulation of the time-dependent profiles.
While, if the source is in motion and its life-time(e.g. $>10^{2}$s) is long
enough that its location (say, $\varphi_{\rm s}$) significantly changes, the
observed time-averaged iron profile should be the integration of the profiles
calculated in this paper along the location of the sources in the motion.
Detailed calculation should be done for the effects of an off-axis source on the
Fe K$\alpha$ line profile and its variation. 
 
Of the future X-ray missions, {\it Astro-E} has the best energy resolution at
the iron K line. {\it XMM} has the largest effective area and it is more
appropriate for detecting the line variability, which can provide the best
quality spectra to examine the disk structure, geometry, ionization state, and
so on (Kunieda, 1996). Recently, Reynolds, Young \& Begelman et al. (1999)
calculated the iron line response to the activation/flaring of a new emitting
region, which may be used to probe the mass
and spin of AGN black holes as well as the X-ray source geometry. Most of their
calculations are performed for the case of one on-axis source. The calculation
for off-axis and moving off-axis flares should be developed further. The
combination of the theoretical calculation and observational data will reveal
important clues of the motion of X-ray sources, the X-ray emission mechanism
and also the underlying physical process.

\section{Conclusion}
In this paper, we explored the effects on time-averaged iron K$\alpha$ line
profiles driven by an off-axis X-ray source. We found different source location
and motion have a significant effect on illumination. The disk region under
the source will receive more flux, which is the most important factor to 
affect the line profiles. We suggest that at least part among the variation of
Fe K$\alpha$ line profiles is caused by the motion of X-ray source.
The time-averaged line profiles can provide information about the 
distribution and motion of X-ray sources.
To reveal more information about the X-ray source nature, time-dependent line profiles
should be explored in future work. 
\section*{Acknowledgments}
We thank the anonymous referee for a critical reading of this paper and many
useful suggestions. Y. Lu thanks the support of Chinese National Science
Foundation and Pan-Deng Project.
%%%%%%%%%%%%%%%%%%%%%%%%%%%%%%%%%%%%%%%%%%%%%%%%%%%%%%%%%%%%%%%%%%%%%%%%%%%%%%

%%%%%%%%%%%%%%%%%%%%%%%%%%%%%%%%%%%%%%%%%%%%%%%%%%%%%%%%%%%%%%%%%%%%%%%%%%%%%%
\appendix
\section[]{Calculation of illumination laws}
The line element in a static axisymmetric space-time can be expressed in this
form (Chandrasekhar 1983):
\begin{equation}
{\rm d}s^{2}=-e^{2\nu}({\rm d}t)^{2}+e^{2\psi}({\rm d}\varphi-\omega{\rm d}t)^{2}+e^{2\mu_{2}}({\rm d}x^{2})
^{2}+e^{2\mu_{3}}({\rm d}x^{3})^{2}.
\end{equation}
A point in that space-time is assigned the four-velocity:
\begin{equation}
u^{0}=\frac{{\rm d}t}{{\rm d}s}=\frac{e^{-\nu}}{\sqrt{1-V^{2}}},
u^{1}=\frac{{\rm d}\varphi}{{\rm d}s} =\varpi u^{0},
u^{\alpha}=\frac{{\rm d}x^{\alpha}}{{\rm d}s}=u^{0}v^{\alpha} (\alpha=2,3),
\end {equation}
where
\begin{equation}
\varpi=\frac{{\rm d}\varphi}{{\rm d}t},v^{\alpha}=\frac{{\rm d}x^{\alpha}}{{\rm d}t} (\alpha=2,3),
\end{equation}
and
\begin{equation}
V^{2}=e^{2\psi-2\nu}(\varpi-\omega)^{2}+e^{2\mu_{2}-2\nu}(v^{2})^{2}+
e^{2\mu_{3}-2\nu}(v^{3})^{2}.
\end{equation}
The line element will degenerate into the Schwarzschild metric if
\begin{equation}
e^{2\nu}=1-2/r,
e^{2\psi}=r^{2}{\rm sin}^{2}\theta,
e^{2\mu_{2}}=\frac{1}{1-2/r},
e^{2\mu_{3}}=r^{2},
{\rm d}x^{2}={\rm d}r,
{\rm d}x^{3}={\rm d}\theta,
\omega=0.
\end{equation}

For a Schwarzschild black hole, we define $R^{\ast}$ as the local frame moving
with the disk materials, $R$ as curvature coordinates frame not moving with
the disk materials and $R'$ as the local rest frame of the X-ray source.
All terms computed in $R^{\ast}$ and $R'$ are labeled with a star or prime
respectively.
In the frame rotating with the disk, including the `k-correction',
the illuminating flux on the disk is given by (Petrucci \& Henri 1997):
\begin{equation}
F_{\rm em}^{\ast}(r_{\rm em},\varphi_{\rm em})=\frac{{\rm d}P_{\rm em}^{\ast}}{{\rm d}S_{\rm em}^{\ast}}
=(\frac{\nu_{\rm em}^{\ast}}{\nu'_{\rm s}})^{2+\alpha}(\frac{{\rm d}P'_{\rm s}}{{\rm d}\Omega'_{\rm s}}
{\rm d}\Omega'_{\rm s})\frac{1}{{\rm d}S_{\rm em}^{\ast}}
\end{equation}
where\\
$\nu'_{\rm s}$ is the photon frequency in the source rest frame;\\
$\nu_{\rm em}^{\ast}$ is the photon frequency in the disk material rest frame;\\
$\alpha$ is the spectral index of the X-ray continuum $P'_{\rm s}(\nu'_{\rm s})\propto
\nu_{\rm s}^{'-\alpha}$ and here we assume $\alpha=1.0$;\\
${\rm d}S_{\rm em}^{\ast}$ is the unit area of the disk in $R^{\ast}$;\\
${\rm d}\Omega'_{\rm s}$ is the unit solid angle in $R'$;\\
$\frac{{\rm d}P'_{\rm s}}{{\rm d}\Omega'_{\rm s}}$ is the emitting power per unit solid angle in
$R'$.

Using the covariance of the space-time quadri-volume between the two inertial 
frames $R$ and $R^{\ast}$, we have
\begin{equation}
{\rm d}S_{\rm em}^{\ast}{\rm d}t_{\rm em}^{\ast}{\rm d}h_{\rm em}^{\ast}={\rm d}S_{\rm em}{\rm d}t_{\rm em}{\rm d}h_{\rm em},
\end{equation}
where ${\rm d}h_{\rm em}^{\ast}$ and ${\rm d}h_{\rm em}$ are the elementary space intervals in the 
direction of the disk axis. Since there is no motion along this axis, 
${\rm d}h_{\rm em}^{\ast}={\rm d}h_{\rm em}$; thus
\begin{eqnarray}
F_{\rm em}^{\ast}(r_{\rm em},\varphi_{\rm em})
=(\frac{{\rm d}P'_{\rm s}}{{\rm d}\Omega'_{\rm s}})\frac{{\rm d}\Omega'_{\rm s}}{{\rm d}S_{\rm em}}
\frac{{\rm d}t_{\rm em}^{\ast}}{{\rm d}t_{\rm em}}(\frac{\nu_{\rm em}^{\ast}}{\nu'_{\rm s}})^{2+\alpha}
\end{eqnarray}
where
\begin{equation}
{\rm d}S_{\rm em}=\frac{r_{\rm em}}{\sqrt{1-2/r_{\rm em}}}{\rm d}r_{\rm em}{\rm d}\varphi_{\rm em};
\end{equation}
\begin{equation}
\frac{{\rm d}t_{\rm em}^{\ast}}{{\rm d}t_{\rm em}}=\sqrt{1-V_{\rm em}^{2}};
\end{equation}
\begin{equation}
\frac{\nu_{\rm em}^{\ast}}{\nu'_{\rm s}}=\frac{\bmath{p_{\rm em}\cdot u_{\rm em}}}
{\bmath{p_{\rm s}\cdot u_{\rm s}}};
\end{equation}
\begin{eqnarray}
\bmath{p_{\rm em}}
&=&(p_{t_{\rm em}}, p_{r_{\rm em}}, p_{\theta_{\rm em}}, p_{\varphi_{\rm em}})\\
&=&(-E, \pm E\frac{\sqrt{r_{\rm em}^{4}-(r_{\rm em}^{2}-2r_{\rm em})(l^{2}+q^{2})}}
{r_{\rm em}^{2}-2r_{\rm em}}, \pm E\sqrt{q^{2}-l^{2}{\rm ctg}^{2}\theta_{\rm em}}, El);
\end{eqnarray}
\begin{equation}
\bmath{u_{\rm em}}=(u_{\rm em}^{t}, u_{\rm em}^{r}, u_{\rm em}^{\theta}, u_{\rm em}^{\varphi})
=(u_{\rm em}^{t}, 0, 0, \varpi_{\rm em}u_{\rm em}^{t});
\end{equation}
\begin{equation}
\varpi_{\rm em}(r_{\rm em})=\frac{1}{r^{3/2}_{\rm em}};
\end{equation}
$\frac{{\rm d}P'_{\rm s}}{{\rm d}\Omega'_{\rm s}}=constant$ because that the emission is assumed to
be isotropic in the source rest frame.

In the X-ray source rest frame $R'$, the local emission angles, expressed in 
spherical polar coordinates, come from projecting $\bmath{p_{\rm s}}$ onto the 
spatial basis vectors:
\begin{equation}
{\rm cos}\theta'_{\rm s}=-\frac{\bmath{p_{\rm s}\cdot e_{\hat{\theta}_{\rm s}}}}
{\bmath{p_{\rm s}\cdot u_{\rm s}}},
{\rm sin}\theta'_{\rm s}{\rm cos}\varphi'_{\rm s}=-\frac{\bmath{p_{\rm s}
\cdot e_{\hat{r}_{\rm s}}}}
{\bmath{p_{\rm s}\cdot u_{\rm s}}},
{\rm sin}\theta'_{\rm s}{\rm sin}\varphi'_{\rm s}=-\frac{\bmath{p_{\rm s}\cdot
e_{\hat{\varphi}_{\rm s}}}}{\bmath{p_{\rm s}\cdot u_{\rm s}}};
\end{equation}
\begin{equation}
\bmath{e_{\hat{\theta}_{\rm s}}}=\frac{1}{r_{\rm s}}\partial_{\theta_{\rm s}},
\bmath{e_{\hat{r}_{\rm s}}}=\sqrt{1-\frac{2}{r_{\rm s}}}\partial_{r_{\rm s}},
\bmath{e_{\hat{\varphi}_{\rm s}}}=\frac{1}{r_{\rm s}{\rm sin}\theta_{\rm s}}\partial_{\varphi_{\rm s}};
\end{equation}
\begin{equation}
\bmath{u_{\rm s}}=(u_{\rm s}^{t}, u_{\rm s}^{r}, u_{\rm s}^{\theta}, u_{\rm s}^{\varphi});
\end{equation}
\begin{eqnarray}
\bmath{p_{\rm s}}&=&(p_{t_{\rm s}}, p_{r_{\rm s}}, p_{\theta_{\rm s}},p_{\varphi_{\rm s}})\\
&=&(-E, \pm E\frac{\sqrt{r_{\rm s}^{4}-(r_{\rm s}^{2}-2r_{\rm s})(l^{2}+q^{2})}}
{r_{\rm s}^{2}-2r_{\rm s}}, \pm E\sqrt{q^{2}-l^{2}{\rm ctg}^{2}\theta_{\rm s}}, El).
\end{eqnarray}

We assume two cases about the motion of the X-ray source:
one is that the source is static $\bmath{u_{\rm s}}=(u_{\rm s}^{t}, 0, 0, 0)$;
the other is that the source has an angular velocity 
$\bmath{u_{\rm s}}=(u_{\rm s}^{t}, 0, 0, \varpi_{\rm s}u_{\rm s}^{t})$
as the disk materials since the corona is generally coupled with the
accretion disk,
where we assume $\varpi_{\rm s}=\frac{1}{2}\varpi_{\rm em}(r_{\rm s})
=\frac{1}{2}\frac{1}{r^{3/2}_{\rm s}}$.
The emission is supposed to be isotropic in the source rest frame and photons
are uniformly produced within the ranges $\theta'_{\rm s}\in[-\pi/2, \pi/2]$
 and $\varphi'_{\rm s} \in[0, 2\pi)$.
For any given ($\theta'_{\rm s}, \varphi'_{\rm s}$), the motion constants
of photons $(l, q)$ can be derived by the above equations (A16). 
Using the photon motion equations (8), we will work out whether the
photons will arrive at the disk surface and the corresponding coordinates
($r_{\rm em}, \frac{\pi}{2}, \varphi_{\rm em}$) can be
calculated. Then, by (A8), the distribution of the illumination
flux (Fig.1)can be estimated.
%%%%%%%%%%%%%%%%%%%%%%%%%%%%%%%%%%%%%%%%%%%%%%%%%%%%%%%%%%%%%%%%%%%%%%%%%%%%%%

\end{document}